\documentclass[prd, amsfonts,nofootinbib, showpacs]{revtex4}
\usepackage{graphicx, epsfig}
\def\beqn{\begin{eqnarray}}
\def\eeqn{\end{eqnarray}}
\relax
\def\ba{\begin{array}}
\def\ea{\end{array}}

\def\beq{\begin{equation}}
\def\eeq{\end{equation}}
\def\bea{\begin{array}}
\def\eea{\end{array}}

\def\to{\rightarrow}

\def\dis{\displaystyle}
\def\f{\frac}
\def\[{\left[}
\def\]{\right]}
\def\({\left(}
\def\){\right)}

\def\qs{{\sqrt{2}}}

\def\Ah{{\widehat{A}}}
\def\ms{{\widetilde{m}}}
\def\sm0{{\widetilde{m}_0}}
\def\sB{{\sin\beta}}

\def\tanB{{\tan\beta}}
\def\cotB{{\cot\beta}}
\def\sM{{\widetilde{\cal M}}}
\def\sQ{{\widetilde{Q}}}
\def\sU{{\widetilde{U}}}
\def\sD{{\widetilde{D}}}

\def\bs{\tilde{b}}
\def\ds{\tilde{d}}
\def\gs{\tilde{g}}
\def\ts{\tilde{t}}
\def\cs{\tilde{c}}

\def\U1em{{U(1)_{\rm em}}}
\def\to{\rightarrow}

\def\sq2{\sqrt{2}}

\def\End{\end{document}}

\def\sm{{\tilde{m}}}
\begin{document}
\title{CKM-suppressed top quark decays $t\to q+W$ in the SM and
  beyond}


\author{J.~L. Diaz-Cruz$^1$, R. Gaitan-Lozano$^2$, G. Lopez  Castro$^3$, C.~E. Pagliarone$^4$}
\vspace{0.6cm}
\affiliation{$^1$ Cuerpo Acad\'emico de Particulas, Campos y Relatividad FCFM, BUAP, Pue., Mexico.}
\affiliation{$^2$ Centro de Investigaciones Te\'oricas, FES-Cuautitlan, UNAM, Mexico.}
\affiliation{$^3$ Departamento de Fisica, Cinvestav, Mexico D.F., Mexico.}
\affiliation{$^4$ Universit\'a degli Studi di Cassino \& INFN Pisa.}
\begin{abstract}
Top quark decays are of particular interest as a mean to
test  the standard model (SM) predictions, both for the
dominant ($t\to b+W$) and rare decays ($t\to q+W, cV, cVV,c\phi^0,bWZ$).
As the latter are highly suppressed, they become an excellent window
to probe the predictions of theories beyond the SM. In particular, we
evaluate the  corrections from new physics to the CKM-suppressed
SM top quark decay $t\to q+W$ ($q=d,s$), both within an effective model with right-handed currents and the MSSM. We also discuss the perspectives to probe those
predictions at the ILC.
\end{abstract}
\maketitle

\section{Introduction.}
 After the discovery of the top quark at Fermilab Tevatron Collider~
\cite{cdf_top,d0_top},
experimental attention has been turned on the examination of its
production mechanisms and decay properties.
Within the Standard Model (SM), the top quark production cross section is
evaluated with an uncertainty of the order of $\sim 15\%$, while it is
assumed to decay to a $W$ boson and a $b$ quark almost $100\%$ of the time.
Due to its exceedingly heavy mass, the top quark is expected to
be somehow related to new physics; it is also considered that the top quark may give some clue
to understand the mechanism of electroweak symmetry breaking.
Thus measuring its properties may serve
as a window for probing physics beyond the SM~\cite{Peccei}.

The interactions of quarks and leptons, with gauge bosons, seem to be correctly
described by the $SU(3)_c\times SU(2)_L\times U(1)_Y$ gauge theory, as
plenty of experimental data shows~\cite{revsm}.
At tree-level, SM neutral interactions are diagonal,
however, flavor changing neutral currents (FCNC) can arise
at loop level.
The fact that FCNC $B$-meson decays have been already detected, at
rates consistent with the SM~\cite{PDG}, represents a great
success for the model itself.
However, SM predictions for top quark related processes
are strongly suppressed, although corresponding
experimental bounds are rather weak.
At the coming LHC it is important to study rare top quark decays, because about
$10^7-10^8$ top pairs will be produced per year.
Thus, rare decays with B.R. of order $10^{-5}-10^{-6}$
may be detectable, depending on the signal.
The presence of any hint for new top quark physics at LHC~\cite{tt-rev},
would motivate further study to clarify the implications of those effects
at the next generation of collider experiments~\cite{tt-LC}.

In this paper we are interested in studying the CKM suppressed decays of the top quark,
 and the perspectives to detect them at the ILC \cite{ckmtop}. It
includes first a short review of top quark decays (section 2), then we
discuss the expected branching ratio for the CKM-suppressed  top quark
decays $t\to q+W$ in the SM (section 3-A); here we also study how the extraction of the CKM
element $V_{ts}$ from B meson data, does depend on the assumptions made and in particular we
determine how the value of $V_{ts}$ changes when one includes a generic right-handed (RH) current,
which could be incorporated using the effective lagrangian approach (section 3-B). This is a
generic approach to new physics that serves to parametrize in a general setting
corrections from new physics to flavor mixing \cite{effmixq}. Then,
 Section 4 includes the predictions for this decay from
the Minimal SUSY extension of the SM (MSSM)~\cite{susyrev}.
 The possibilities to detect such modes at the future  International Linear
 Collider (ILC) is presented in section 5, while the conclusions and a
 summary of results are shown in section 6.

\section{A survey of top decays in the SM and beyond}

Because of the structure of the SM, the $W$ boson coupling to
fermion pairs ($td_iW^\pm$), is proportional to the CKM element
$V_{td_i}$. Thus the decay $t\to b+W$ dominates its BR's.
Radiative corrections to this mode have been evaluated in the
literature, both in the SM and some extensions, mainly within the
minimal SUSY extension of the SM (MSSM). In general, such corrections
are at most of order $10\%$, and therefore difficult to detect at
hadron colliders, but may be at the reach of the ILC.

FCNC top quark decays, such as $t\to c \gamma$, $t\to c g$, $t\to c
Z$ and $t\to c \phi$ have been studied, for some time, in the
context of both the SM and  new physics \cite{bmele-rev}. In the SM,
the branching ratio of FCNC top decays is extremely suppressed, as
it is summarized in table I. The rare top quark decay $t\to
c+\gamma$ was calculated first in ref. \cite{ourtcg} in the SM and
some extensions, the result implied a suppressed B.R., less than
about $10^{-10}$, which was confirmed when subsequent analysis
\cite{nexttcg} that included the correct top mass value and gave
$BR(t \to c+\gamma)= 5 \times 10^{-13}$. The decays  $t\to c+Z$ and
$t\to c+g$ were also calculated in refs. \cite{nexttcg}. The
resulting branching ratios turned out to be $BR(t\to c+Z)=1.3 \times
10^{-13}$ and $BR(t\to c+g)=5 \times10^{-11}$. None of them seem
detectable at LHC nor at the ILC.

The top-charm coupling with the SM Higgs  $\phi^0$ could also be
induced at one-loop level \cite{SM-tch}.
The resulting  branching ratio is given by $BR(t\to c+\phi^0)=10^{-15}$ ,
which does not seem detectable neither.
The FCNC top decays involving a pair of vector bosons in the final
state, $t \to cVV$, can also be of interest \cite{ourtcvv}. Although one
could expect such
modes to be even more suppressed than the ones with a single vector boson,
the appearance of an intermediate scalar resonance, as in the previous
case, could enhance the B.R.. Furthermore,  because of the large top quark
mass, it also seems possible to allow the tree-level decay  $t\to b+WZ$,
at least close to threshold.

\bigskip

On the other hand, the top decay into the light quarks $t\to W
+d(s)$ is suppressed, as they are proportional to $V_{td(s)}$.
Furthermore, it is unlikely that these modes could be detected at
all at hadron colliders. Probably for this reason, the SM
corrections to this mode have not been studied, though the QCD corrections should
be the similar for both modes. However, in
extensions of the SM, it may be possible to get a large enhancement
that could even make it detectable at the ILC, as will be shown in
the next section.

Some typical results for the top decays in the SM are summarized in Table
I. This table also includes, for comparison, the results for top
branching ratios from models beyond the SM, in particular from the
THDM-III \cite{roberto} and SUSY \cite{diaz}, which will be discussed in what follows.

\begin{table}[t!]
\begin{center}
\begin{tabular}{| c| c| c| c| }
\hline\hline
{\bf BR\ } & SM & THDM-III & MSSM \\
\hline
$\mathbf{BR(t \to sW)}$ & $2.2 \times 10^{-3}$ & $\sim 10^{-3}$ &
$10^{-3}-10^{-2}$ \\
\hline
$\mathbf{BR(t \to c\phi^0)}$ & $10^{-13}-10^{-15}$ & $\sim 10^{-2}$ &
$10^{-5}-10^{-4}$\\
\hline
$\mathbf{BR(t \to c \gamma)}$ & $5\times 10^{-13}$ & $< 10^{-6}$ &  $<10^{-7}$ \\
\hline
$\mathbf{BR(t \to c Z)}$ & $1.3 \times 10^{-13}$ &  $< 10^{-6}$ & $<10^{-7}$ \\
\hline
$\mathbf{BR(t \to c g)}$ & $5\times 10^{-11}$ & $< 10^{-6}$ & $< 10^{-5}$ \\
\hline
$\mathbf{BR(t \to c \gamma\gamma)}$ & $<10^{-16}$ & $\sim 10^{-4}$ &  $<10^{-8}$ \\
\hline
$\mathbf{BR(t \to c WW)}$ & $2\times 10^{-13}$  & $10^{-4}-10^{-3}$ &
??  \\
\hline
$\mathbf{BR(t \to cZZ)}$ & -- & $10^{-5}-10^{-3}$ & ?? \\
\hline
$\mathbf{BR(t \to bWZ)}$ & $2\times 10^{-6}$ & $\simeq 10^{-4}$ & ?? \\
\hline
\end{tabular}
\end{center}
\vspace{-0.55cm}
\caption{ Branching ratios for some CKM-suppressed and FCNC top quark
decays in the SM and beyond, for $m_t=173.5-178$ GeV. Decays into a pair of
massive gauge bosons include  finite width effects of final state unstable
particles \cite{fwe}.}
\vspace{-0.4cm}
\end{table}

\section{The top quark decays $t\to q+W$ in the SM and
with RH currents: CKM analysis}
 Although the decay $t\to s+W$ is expected to be suppressed in the SM, it is
possible that new physics (e.g. LR models or SUSY) could induce an
enhancement on this mode that could make it to be at the reach of ILC.
This will have the attractive of allowing a tree-level determination
of the CKM element  $V_{ts}$, which otherwise need to be extracted
from B-physics using one-loop induced processes, as we discuss next.

\subsection{The decay $t\to s+W$ with standard CKM}
 Within the SM the rate of $t\to q+W$ decays at the tree-level is given by
:
\begin{equation}
\Gamma^{SM} (t\to qW)= \frac{G_Fm_t^3}{8\pi \sqrt{2}}|V_{tq}|^2 \left\{
x^2(1-2x^2+y^2)+(1-y^2)^2\right\}  \nonumber  \\
   \times \lambda^{1/2}(1,x^2,y^2)\ ,
\end{equation}
where: $x=m_W/m_t,\ y=m_q/m_t$ ($q=d,s,b$). The final quark masses can be
safely neglected in  all the relevant cases (even for $q=b$,
$y^2=(m_b/m_t)^2=8.4 \times
10^{-4}$ is negligible). Thus, the decay is proportional to the CKM
elements $V_{tq}$, which are not so well known at present.

  Here and thereafter we will assume $|V_{tb}|=1$. Measurements of $b\to
s\gamma$ by CLEO, BABAR and
Belle, and its description in the SM framework  give the following
constraint on $|V_{ts}|$ \cite{PDG}:
\begin{equation}
|V_{ts}| = (40.6 \pm 2.7) \times 10^{-3}  \ .
\end{equation}
Assuming also top quark dominance of box diagrams contributions to
$\Delta M_d$ in $B_d-\bar{B}_d$ mixing yields the following SM constraint
\cite{PDG}:
\begin{equation}
|V_{td}| = (7.4\pm  0.8) \times 10^{-3} \ .
\end{equation}

   The corresponding branching fractions for the
CKM-suppressed $t\to q'W$  ($q'=s,\ d$)  decays in the SM are:
\begin{eqnarray}
B(t\to q'W) &=& \frac{\Gamma( t\to q'W)}{\sum_q \Gamma(t\to qW)} \nonumber
\\
&\approx & |V_{tq'}|^2 \nonumber \\
& = &
\left\{ \begin{array}{l} 1.65 \times 10^{-3}, \  \mbox{\rm for}  \  q'=s\
,
\\ 5.5 \times 10^{-5}, \ \mbox{\rm for} \ q'=d \ .  \end{array} \right.
\end{eqnarray}

   Note that, in the case that the charged fermion couplings are
modified by new physics, for instance as in the effective lagrangian
approach \cite{effmixq} or the MSSM, to be discussed next,
the values given in eq. (2,3) may not hold anymore. In particular, the
dominance of the top quark in loop contributions to $b \to s\gamma$ and
$\Delta M_{B_d}$ can be spoiled by the non-unitarity of the quark
mixing matrix (as in ref. \cite{effmixq}) or other new physics effects.
Note also that other indirect constraints on $V_{ts},\ V_{td}$ can be
obtained from the rare $K^+ \to \pi^+ \nu \bar{\nu}$ decay, the ratio
of mass differences of neutral B mesons, and from the exclusive  $B\to
\rho(K^*)\gamma$ decays. At present, theoretical and/or experimental
uncertainties on these observables make them less competitive
sources of information \cite{PDG}.

\subsection{The decay $t \rightarrow W + s$ with RH top quark couplings}
In this section we give an estimate of the width of top decay $t
\rightarrow s + W$ that stems from the $tsW$ coupling extracted from $b\to
s \gamma$ decays, when a Left-Right modification is allowed. Namely,
here we also study how the extraction of the CKM element $V_{ts}$ from B meson data,
does depend on the assumptions made and in particular we determine how the value of
$V_{ts}$ changes when one includes a generic right-handed (RH) current,
which could be incorporated using the effective lagrangian approach.

We start from the analysis of the decay $B \rightarrow X_{s}
\gamma$ as done by the authors in \cite{kaganneubert} who define the
following ratios of Wilson coefficients ($C_7,C_8$) associated with the
dipole operators $O_7$ and $O_8$,

\begin{equation}
r_7 = \frac{C_7(m_W)}{C_7^{SM}(m_W)}\ ,
\end{equation}

\begin{equation}
r_8 = \frac{C_8(m_W)}{C_8^{SM}(m_W)}\ ,
\end{equation}

\begin{equation}
r_7^R = \frac{C_7^R(m_W)}{C_7^{SM}(m_W)}\ ,
\end{equation}

\begin{equation}
r_8^R = \frac{C_8^R(m_W)}{C_8^{SM}(m_W)} \ ,
\end{equation}
while $C^{R}_{7,8}$ denote the Wilson coefficients of new dipole
operators with opposite chirality to that of the Standard Model, and
are evaluated at the scale $m_W$,

Then, we define the following useful ratio for the inclusive
$B \rightarrow X_{s} \gamma$ decay,
\begin{eqnarray}
\frac{1}{N_{SL}}B(B \rightarrow X_{s}
\gamma)|_{E_{\gamma}>(1 - \delta)E_{\gamma}^{max}}\nonumber
          &=&
B_{22}(\delta) + B_{77}(\delta)(|r_7|^2 + |r_7^R|^2) +
B_{88}(\delta)(|r_8|^2 + |r_8^R|^2) \\
& &  + B_{27}(\delta)Re(r_7) +
B_{28}(\delta)Re(r_8)\nonumber \\
&& + B_{78}(\delta)[Re(r_7 r_8^{*}) + Re(r_7^R r_8^{R*})]\ . \nonumber
\end{eqnarray}
where $N_{SL} = B(B\rightarrow X_c e \bar{\nu})/0.105$ is a normalization
factor to be determined from experiment.

The coefficients $B_{ij}$ depend on the kinematical cut $\delta$, with
numerical values given in Ref. \cite{kaganneubert}. The $\delta$
parameter is defined by the condition that the photon energy be
above a threshold given by $E_{\gamma}>E_0=(1 -
\delta)E_{\gamma}^{max}$, where $E_{\gamma}^{max} = m_b/2$ is
the maximum photon energy attainable in the parton model.

Next, considering the contributions from the Standard Model and from
New Physics (RH couplings), we can write:
\begin{equation}
B(B \rightarrow X_s \gamma) = B^{SM}(B \rightarrow X_s \gamma) +
B^{NP}(B \rightarrow X_s \gamma)\ .
\end{equation}

In the case of the SM contributions we use the results of Neubert
\cite{neubert} which include the complete NNLO effects. For the NP
contributions we choose the calculations of  Kagan and Neubert
\cite{kaganneubert} given above.
If we identify $B(B\rightarrow X_s\gamma)$ with the experimental
result, we can write the previous equation as follows:

\begin{equation}
\label{btot} B^{exp}(\delta) = |V^*_{ts}V_{tb}|^2\left[
\frac{B^{SM}(B\rightarrow
X_s\gamma)|_{E_0=1.8\ GeV}}{(0.0404^{+0.0016}_{-0.0006})^2} +
\frac{X_{LR}}{(0.953\pm
0.0195)^2}\cdot \frac{B(B \rightarrow X_c
l \nu)}{0.105|V_{cb}|^2} \right]
\end{equation}
where

\begin{equation}
\label{lin} X_{LR} = B_{77}(\delta)|r_7^R|^2 +
B_{88}(\delta)|r_8^R|^2 + B_{78}(\delta)Re(r_7^R r_8^{R*})\ ,
\end{equation}
and the SM prediction reads \cite{neubert}
\begin{equation}
B^{SM}(B\to X_s\gamma)|_{E_0=1.8\ GeV}= (3.38^{+0.31}_{-0.42}\pm
0.31)\times 10^{-4}\ .
\end{equation}

In order to evaluate numerically the Wilson coefficients
that enter in  the definition of $X_{LR}$, eq.(~\ref{lin}), we use the
following expressions \cite{lee}: $C_7^{SM}(m_W) = F(x_t)$, $C_8^{SM}(m_W)
= G(x_t)$,
$C_7^R(m_W) = \xi (\frac{m_t}{m_b}) \bar{F}(x_t)$, and $C_8^R(m_W) =
\xi (\frac{m_t}{m_b}) \bar{G}(x_t)$, where $F(x_t)$ and $G(x_t)$ are the
Inami-Lim loop functions \cite{inami} and $\xi$ denotes the mixing angle
between the charged gauge bosons $W_L$ and $W_R$ of the left-right
symmetric model. In the evaluation of the Inami-Lim functions we use
$m_t=174$ GeV.

Now, we consider the most recent experimental measurement of $B\rightarrow
X_s\gamma$ reported by Belle (see for example, Ref. \cite{Belle}):

\begin{equation} B^{exp}(\bar{B} \rightarrow X_s
\gamma)|_{E_0=1.8\ GeV}  = (3.38\pm 0.30  \pm 0.29) \times 10^{-4}\ .
\end{equation}
We can easily check that if NP contributions in eq. (\ref{btot}) are
turned off, we recover the value of $|V_{ts}|$ given in eq. (2).

Now, we can insert the experimental value of $B(B\to X_s\gamma)$ into
eq. (\ref{btot}) and derive the modified values of $|V_{ts}|$ by allowing
the parameter $\xi$ to vary ($|\xi| \leq 0.05$). We will use $|V_{cb}| =
41.6 \times 10^{-3}$ \cite{PDG} and $B(B\to X_cl\nu)=(10.75\pm 0.16)\%$
\cite{hfag}. The corresponding results are shown in Table II. In the same
Table we also display the values of the $t\to sW$ branching ratios in the
extension of the SM, with RH currents which were calculated according to the expression
\begin{equation}
B(t \rightarrow s + W) =|V_{ts}|^2(1 + \xi^2)\ .
\end{equation}
Thus, we can conclude that the overall effect of the LR mixing angle
$\xi$, is to decrease the prediction for $B(t \rightarrow s + W)$.
This effect is sizeable and can eventually be discriminated at
linear colliders. Conversely, if the $B(t \rightarrow s + W)$ is found to
be in  close agreement with the SM prediction, this would manifest into a
very strong constraint on the LR mixing angle.

\vspace{.3cm}

\begin{table}
\begin{tabular}{|c|c|c|}\hline
$|\xi|$ &$|V_{ts}|$ &$B(t \rightarrow s + W)$\\
\hline
0 & $40.4\times 10^{-3}$ & $1.63\times 10^{-3}$ \\
0.01 & $34.9\times 10^{-3}$ & $1.22 \times 10^{-3}$ \\
0.02 & $26.3\times 10^{-3}$ & $0.69 \times 10^{-3}$ \\
0.03 & $20.0\times 10^{-3}$ & $0.40\times 10^{-3}$ \\
0.04 & $15.9 \times 10^{-3}$ & $0.25\times 10^{-3}$ \\
0.05 & $13.1\times 10^{-3}$ & $0.17\times 10^{-3}$ \\
\hline
\end{tabular}
\caption{Branching fractions for $t \to sW$ in the LR model by \\ using
the experimental $B\to X_s \gamma$ constraints.}
\end{table}

\section{Top quarks decays $t\to q+W$ in the MSSM}

We shall discuss now top decays in the context of Supersymmetry (SUSY),
which is
studied mainly in connection with a possible solution of the hierarchy
problem~\cite{susyrev}.  The Minimal Supersymmetric extension of the SM (MSSM),
has well known attractive features, such as allowing gauge coupling
unification and radiative electroweak symmetry breaking (EWSB).
In addition  SUSY also predicts tau-bottom Yukawa unification
and provides a dark matter candidate.
Within the MSSM, there are new sources of FCNC, and it may be difficult
to satisfy current bounds on such transitions in a general SUSY
breaking scheme.
 Generic entries for sfermion mass matrices produce so
large FCNC that some mechanism must be found to suppress them,
which is known as the SUSY flavour problem.
Best known solutions to this problem include:
{\it{i) Degeneracy, ii) Decoupling and iii) Alignment }}.
In particular, the interactions of the gluino with squarks and quarks,
could involve different families, which in turn could mediate FCNC
transitions for quarks. Such gluino interactions can produce large
enhancements on the top quark FCNC decays.
In this section we shall discuss a particular ansatz that
satisfies current FCNC constraints, and still allows a large
stop-scharm mixing, which in turn could induce important
corrections to the rare top quark decays.

\subsection{The MSSM with large Stop-scharm mixing.}
In the following we shall review the formalism presented in
ref. \cite{ourtchsup}, which shall be applied to evaluate the
top decay $t\to s+W$, which has a larger branching ratio than
$t\to d+W$.

Within the MSSM quark masses, as well as SUSY conserving F- terms,
are contained in the superpotential of the model,
\beq
W= \lambda^u Q H_u U + \lambda^d Q H_d D
\eeq
where $\lambda^{u,d}$ denote the $3\times 3$ Yukawa
matrices; $Q,U,D$ are the superfields that denote the quark/squark
doublets and singlets.

The soft-breaking squark-sector contains the following
quadratic mass and trilinear $A$-terms,
\beq
\bea{l}
- {\sQ^{\dag}_i (M_{\sQ}^{2})_{ij}\sQ_j
- \sU^{\dag}_i (M_{\sU}^{2})_{ij}\sU_j
- \sD^{\dag}_i (M_{\sD}^{2})_{ij}\sD_j } \\[2mm]
+ ( A_{u}^{ij}\sQ_i H_u\sU_j
  -A_{d}^{ij}\sQ_i H_d\sD_j + {\rm c.c.} )\,,
\eea
\label{eq:A-term}
\eeq
with $M_{\sQ,\sU,\sD}^2$ and
$A_{u,d}$ being $3\times3$ matrices in squark flavor-space.
Here $\sQ_i, \sU_i, \sD_i$ denote the squark doublet and singlets
for family $i$.
This gives a generic $6\times6$ mass matrix,
\beq
\sM^2_u =\left\lgroup
     \bea{ll}
          M_{LL}^2         &  M_{LR}^2\\[1.5mm]
          M_{LR}^{2\,\dag}   &  M_{RR}^2
     \eea
         \right\rgroup ,
\label{eq:MU6x6}
\eeq
in the up-squark sector, where

\beq
\bea{ll}
M_{LL}^2 &= M_{\sQ}^2+M_u^2+\f{1}{6}\cos2\beta \,(4m_w^2-m_z^2)\,, \\[2mm]
M_{RR}^2 &= M_{\sU}^2+M_u^2+\f{2}{3}\cos2\beta\sin^2\theta_w\, m_z^2\,,  \\[1mm]
M_{LR}^2 &= \dis A_u v\,\sB/\sqrt{2}-M_u\,\mu\,\cotB \,,
\eea
\label{eq:MU3x3}
\eeq
with $m_{w,z}$ the masses of $(W^\pm,\,Z^0)$ and $M_u$ the up-quark
mass matrix. For convenience, we will choose hereafter the
super Cabibbo-Kobayashi-Maskawa (CKM) basis for squarks so that
in (\ref{eq:MU3x3}),
$A_u \to A_u' = K_{UL} A_u K_{UR}^\dag$ and
$M_u \to M_u^{\rm diag}$, etc, with $K_{UL,R}$ denoting the rotation matrices
for $M_u$ diagonalization.

In our {\it minimal} Type-A scheme, we consider all large FCNCs to
{\it solely} come from non-diagonal $A_u'$ in the up-sector, and those in
the down-sector to be negligible,
i.e. we define at the weak scale,
\vspace*{-2mm}
\beq
A_u' =
      \left\lgroup
      \bea{ccc}
      0 & 0 & 0\\
      0 & 0 & x\\
      0 & y & 1
      \eea
      \right\rgroup A \,,
\label{eq:Au}
\eeq
where $x$ and $y$ can be of $ O(1)$, representing a naturally
large flavor-mixing associated with $\ts - \cs$ sector.
 Such a minimal scheme of FCNC is compelling as it is consistent
with all experimental data as well as the
theoretical CCB/VS bounds \cite{casas}.
 Similar pattern may be also defined for $A_d$ in the down sector,
but the strong  CCB/VS bounds permit $O(1)$ $\bs-\ds$ mixing
only for very large $\tan\beta$, because $m_b << m_t$.
Thus to allow a full range of $\tan\beta$ we consider an
almost diagonal $A_d$. Moreover, identifying the
non-diagonal $A_u$ as the {\it only}
source of observable FCNC phenomena for Type-A schemes implies
that the squark-mass-matrices $M_{\sQ,\sU}^2$ in
eqs.\,(\ref{eq:MU6x6})-(\ref{eq:MU3x3}) to be nearly diagonal.
 For simplicity we define
\beq
M_{LL}^2 \,\simeq\, M_{RR}^2\, \simeq\,\sm0^2\,{\bf I}_{3\times3}\,,
\label{eq:Degen}
\eeq
with $\sm0$ a common scale of scalar-mass.

Within this minimal Type-A scheme, we observe that the first family squarks
$\tilde{u}_{L,R}$ decouple from the rest in (\ref{eq:MU6x6}) so
that this $6\times6$ mass-matrix is reduced to a $4\times4$ matrix,
\beq
\sM_{ct}^2 =
      \left\lgroup
      \bea{cccc}
      ~{\ms}_{0}^2~  &   0             &   0            &   A_x\\[1mm]
      0              &   ~\ms_{0}^2~   &   A_y~         &   0  \\[1mm]
      0              &   A_y~          &  ~\ms_{0}^2~   &  -X_t\\[1mm]
      A_x~           &   0             &   -X_t~        &  ~\ms_{0}^2
      \eea
      \right\rgroup
\label{eq:Mct4x4}
\eeq
for squarks $(\cs_L,\,\cs_R,\,\ts_L,\,\ts_R)$, where
\beq
\bea{l}
A_x = x\Ah,~~A_y = y\Ah,~~\Ah = Av\,\sB/\sqrt{2}\,,  \\[1mm]
X_t = \Ah - \mu\,m_t\,\cotB \,.                     \\ 
\eea
\label{eq:MctADD}
\eeq
In (\ref{eq:Mct4x4}), we ignore terms suppressed by tiny factors of
$O(m_c)$ or
smaller. The reduced squark mass matrix (\ref{eq:Mct4x4}) has
6 zero-entries in total and is simple enough to allow an exact
diagonalization.
Below, we summarize the general diagonalization of $4\times4$ matrix
(\ref{eq:Mct4x4}) for any $(x,\,y)$.
The mass-eigenvalues of
the eigenstates  $(\cs_1,\,\cs_2,\,\ts_1,\,\ts_2)$ are,
\beq
\bea{ll}
M_{\cs1,2}^2 & = \sm0^2 \mp\f{1}{2}|\sqrt{\omega_+}-\sqrt{\omega_-}|\,, \\[2mm]
M_{\ts1,2}^2 & = \sm0^2 \mp\f{1}{2}|\sqrt{\omega_+}+\sqrt{\omega_-}|\,,
\eea
\label{eq:Mass}
\eeq
where $~\omega_\pm = X_t^2+(A_x\pm A_y)^2\,$.
From (\ref{eq:Mass}), we can deduce the mass-spectrum
of stop-scharm sector,
\beq
M_{\ts1} < M_{\cs1} < M_{\cs2} < M_{\ts2} \,.
\label{eq:Mspectrum}
\eeq
In Eq. (\ref{eq:Mass}), the stop $\ts_1$ can be
as light as $120-300$\,GeV for the typical range of
$\sm0^2 \simeq 0.5-1$\,TeV.
Then, the $4\times4$ rotation matrix of the diagonalization can be derived,
\beq
\bea{l}
\left\lgroup
\bea{l}
\cs_L\\
\cs_R\\
\ts_L\\
\ts_R
\eea
\right\rgroup
\!\!=\!\!
\left\lgroup
\bea{rrrr}
 c_1c_3  &  c_1s_3  & s_1s_4  &  s_1c_4  \\
-c_2s_3  &  c_2c_3  & s_2c_4  & -s_2s_4  \\
-s_1c_3  & -s_1s_3  & c_1s_4  &  c_1c_4  \\
 s_2s_3  & -s_2c_3  & c_2c_4  & -c_2s_4
\eea
\right\rgroup
\!\!
\left\lgroup
\bea{l}
\cs_1\\
\cs_2\\
\ts_1\\
\ts_2
\eea
\right\rgroup , \\[8mm]
s_{1,2}=\dis \f{1}{\qs}
\[1-\f{X_t^2\mp A_x^2\pm A_y^2}{\sqrt{~\omega_+\omega_-}} \]^{1/2}\!,~~~
s_4
=\f{1}{\qs}\,,
\eea
\label{eq:rotation}
\eeq
and $s_3=0$ (if $xy=0$), or, $s_3=1/\qs$
(if $xy\neq 0$), where $s_j^2+c_j^2=1$,
and $s_i=\sin \theta_i,\, c_i=\cos\theta_i$.
The rotation (\ref{eq:rotation}) allows us to derive all relevant
new Feynman rules in mass-eigenstates without relying on
``mass-insertions''.

Similar pattern may be also
defined for $A_d$ in the down-sector, but $O(1)$ mixings between $\tilde{b}$
and $\tilde{s}$ are allowed only for very large $\tanB$ by the strong
CCB and VS bounds since $m_b \ll m_t$. To allow full range of
$\tanB$ and focus our minimal scheme on the up-sector, we consider an almost
diagonal $A_d$.  Furthermore, since  the large hierarchy
$A_{33} >> A_{22} >> A_{11}$, is also enforced by CCB and
VS bounds, we shall take  $A_{33}=A_b \neq 0$ and $A_{11}=A_{22}=0$,
together with mass-degeneracy for the soft-breaking
mass matrices as required by FCNC bounds. Then squark
mixing only occurs for the third family,
i.e. for $(\bs_L,\,\bs_R )$,
which is described by the following $2\times 2$ mass matrix,
\beq
\sM_{d}^2 =
      \left\lgroup
      \bea{cc}
  ~\ms_{0}^2~   &  -X_b\\[1mm]
    -X_b~        &  ~\ms_{0}^2
      \eea
      \right\rgroup
\label{eq:Mb2x2}
\eeq
 where
$X_b = -(A_b + \mu \tanB) m_b $. The mass eigenvalues for
the eigenstates  $(\bs_1,\,\bs_2\,)$ are,
\beq
M_{\bs1,2}^2  = \sm0^2 +\f{1}{2}[\hat{\Delta}_b  \mp
                \sqrt{\Delta_b+4X_b^2 }  ]\,,
\eeq
where
$\Delta_b =  \f{\cos 2\beta}{6}
                    [2m^2_w+(1-2s^2_W)m^2_Z] \,$ and
$\hat{\Delta}_b = - \f{\cos 2\beta}{6}
                    [2m^2_w+(1+s^2_W)m^2_Z] \,$,
and the mixing angle of the 2x2 diagonalization is given by:
\beq
s_b(c_b) = \frac{1}{\sqrt{2}} [ 1\mp
    \frac{\Delta_b}{\sqrt{\Delta^2_b+4X_b }  } ]^{1/2}.
\eeq

\subsection{SUSY corrections to the decay $t\to s+W$}

 Within the MSSM, the corrections to the dominant decay mode $t\to bW$
are of the order $1-10 \%$; since $\Gamma(t\to b+W)\simeq 1$ GeV, then one
has that the size of SUSY corrections are of the order $.01-.1$ GeV.
Now, one can estimate the SUSY corrections to the decay $t\to s+W$
by including an extra mixing angle $sin\theta_{ts}$ (among stop and
scharm, which will appear in the loops with gluino-squarks in the
internal lines) , this mixing angle  could be of order $.1-.5$.
Therefore the SUSY correction to the decay $t\to s+W$, would be in the
range $10^{-4}- 10^{-2}$.
Since  $\Gamma(t\to s+W) \simeq 1.3\times 10^{-3}$ GeV, and one
has that the corrections could even be of order $100 \%$,
which can help to make it detectable at ILC.

We shall now present a detailed loop calculation within the context of
the particular mixing scheme, where a large stop-scharm mixing is
allowed, that was discussed in the previous subsection.
The full radiatively corrected coupling $tqW$ can be written as follows,
\vspace*{-1.5mm}
\beq
\bea{rl}
\dis
{\Gamma}^\mu_{tsw} & =~
 -\frac{g_2}{\sqrt{2}}\gamma^\mu P_L  K^{eff}_{ts} \,,\\[1mm]
\dis
 K^{eff}_{ts} & =~ K^{0}_{ts} + \Delta K^{V}_{ts} + \Delta K^{S}_{ts} \,,
\eea
\label{eq:tsW}
\eeq
which includes the tree level vertex, which is proportional to the CKM
element ($K^0_{ts}$), the
corrections from stop-scharm-gluino triangle loops ($\Delta K^V_{ts}$),
as well as the stop- and scharm-gluino self-energy loops  ($\Delta K^S_{ts}$).

The result for the one-loop vertex corrections in Type-A model (with
$x=y$) is
\vspace*{-1.5mm}
\beq
\Delta K^{V}_{ts}  =~ \frac{\alpha_s K^0_{22} }{6 \pi} (4 \sin \theta_1 \cos \theta_1)
                [ -C_{24}(M^2_{\gs},  M^2_{\cs1}, M^2_{\bs} ) - C_{24}(M^2_{\gs},  M^2_{\cs2}, M^2_{\bs} )
                  +C_{24}(M^2_{\gs},  M^2_{\ts1}, M^2_{\bs} ) +
          C_{24}(M^2_{\gs},  M^2_{\ts2}, M^2_{\bs} ) ]
\eeq where $C_{24}$ are the 3-point $C$-function of
Passarino-Veltman (PV), which depends on the gluino and squark
masses. The Sin and Cos of the Mixing angle $\theta_1$ are given in
equation 16.

The dominant correction arising from the self-energy graphs
(with $x=y$) can be expressed as:
\vspace*{-1.5mm}
\beq
\Delta K^{S}_{ts}  =~ \frac{\alpha_s K^0_{22} }{6 \pi} (\sin \theta_1 \cos \theta_1)
                [ -B_{1}(M^2_{\gs},  M^2_{\cs1} ) - B_{1}(M^2_{\gs} , M^2_{\cs2})
                  +B_{1}(M^2_{\gs},  M^2_{\ts1} ) + B_{1}(M^2_{\gs},  M^2_{\ts2} ) ]
\eeq
where
$B_{1}$ denotes the 2-point $C$-function of Passarino-Veltman (PV), which
depends on the gluino and squark masses.

The results for the effective vertex $K^{eff}_{ts}$ are shown in
Table/Figure 2 for $X = 0.9$, $m_{0} = 500$ GeV and several values
of $\tan\beta$ and gluino mass. We notice that the correction can be
larger than the tree-level result by about one order of magnitude. Thus, one could
have $BR(t \to s W)$ even of order $10^{-1}$, which looks promising to be detectable at
the future ILC.
\begin{table}[t!]
\begin{center}
\begin{tabular}{| c| c| c| c| c| }
\hline\hline
$\tan\beta$ & $M_{\gs}$ & $K^{eff}_{32}$ & $r= K^{eff}_{32}/K^0_{32}$
&$B(t \rightarrow s + W)$\\
\hline
 5  & 300  & 0.043 & 1.14  & 5.06 $\times 10^{-3}$\\
\hline
    & 400  & 0.15  & 4.02  & 3.78 $\times 10^{-2}$\\
\hline
    & 500  & 0.79 &  21.27 & 9.98 $\times 10^{-1}$\\
\hline
    & 1000 & 0.067  & 1.79 & 9.25 $\times 10^{-3}$\\
\hline
  20  & 300 & 0.046  & 1.23 & 5.53 $\times 10^{-3}$\\
\hline
    &  400 & 0.39  & 10.37  & 2.39 $\times 10^{-1}$\\
\hline
    &  500 & 0.23  & 6.37  & 9.15 $\times 10^{-2}$\\
\hline
    &  1000 & 0.076  & 2.05 & 8.19 $\times 10^{-2}$\\
\hline
  50  & 300 & 0.047 & 1.26 & 5.69 $\times 10^{-3}$\\
\hline
    &  400 & 0.50 & 13.41 & 3.98 $\times 10^{-1}$\\
\hline
    &  500 & 0.19 & 5.12 & 5.99 $\times 10^{-2}$\\
\hline
    & 1000 & 0.078 & 2.10  & 1.19 $\times 10^{-2}$\\
\hline
\end{tabular}
\end{center}
\vspace{-0.55cm}
\caption{SUSY corrections to the CKM-suppressed top vertex $tsW$. }
\vspace{-0.4cm}
\end{table}

\section{Detection of top decays at the ILC}

Direct measurement of $|V_{td}|$ or $|V_{ts}|$ are not possible at LHC. We will explore
then the possibility of detecting top transition to light quarks with the next generation
lepton collider experiments.

\noindent The International Linear Collider (ILC) is a proposed electron-positron collider
whose design is being addressed in the context of the Global Design Effort~\cite{GDR}.
ILC has been agreed in a world-wide consensus to be the next large high energy
physics experimental facility.
The nature of the ILC's electron-positron collisions would give it the capability
to answer compelling questions that, discoveries at the LHC, will raise from the
identity of dark matter to the existence of extra dimensions.
The overall system design~\cite{ILC} has been chosen to realize the physics requirements with a maximum
center of mass energy of 500 GeV and a peak luminosity of  $2 \times 10^{34}cm^{-2}s^{-1}$.
The total footprint is around $31$ km;
two facing linear accelerators, main linacs, will accelerate electron and positron
beams from their injected energy of $15$ GeV to the final beam energy of $250$ GeV or less,
over a combined lenght of $23$ km.
The electron source, the damping rings, and the positron
auxiliary (`keep-alive') source are centrally located around the interaction region (IR). The
plane of the damping rings is elevated by $10$ m above that of the beam delivery system to
avoid interference.
The baseline configuration furthermore foresees the possibility of an upgrade to energies of about
1 TeV. In order to upgrade the machine to $E_{cms}$= $1$ TeV, the linacs and the beam transport lines from
the damping rings would be extended by another $11$ km each.
While in hadron collisions, it is technically feasible to reach the highest centre-of-mass energies,
often usefull in order to discover new particles,
in $e^+ e^-$ collisions the highest precision for a measurement can
be achieved. High-precision physics at the ILC is made possible in particular by the collision of point-like objects
with exactly defined initial conditions, by the tunable collision energy and by the possibility of working with
polarized  beams.

The present phenomenological work have been performed
assuming a center of mass energy of $\sqrt s=$ $500$ GeV and an integrated luminosity,
considering two running experiments, taking data at the same time, for about 4 years, plus the year $0$,
with a total integrated luminosity of $1$ ab$^{-1}$~\cite{ourwork}.
Events are generated using the PANDORA-PYTHIA Montecarlo generator and applying a parametric
detector simulation.
Heavy quarks, $b-$jets and $c-$jets, are tagged using their well known unambiguous properties such as their
mass and their long lifetimes.
To tag light-quark jets is much more difficult but anyhow this is needed in
order to get meaningful measurement of the CKM matrix elements.
The technique used, in the present work, is the so called
large flavour tagging method (LFTM)~\cite{Letts-Mattig}.
Particles, with large fraction $x_p=$ $2p/E_{cm}$ of the momentum, carry information about
their primary flavour. Then it is possible to define a class of functions:
$\eta^ì_q(x^ì_p)$ that represent the  probability, for a quark of a flavour $q$,
to develop into a jet in which $i$ is the particle that brings the largest $x_p=$ $2p/E_{cm}$.
Tagging efficiencies can be extracted with almost no reliance on the hadronization model,
using a sample of $Z^0$, by evaluating single and double tag probability functions.
In order to simplify the calculations, hadronisation symmetries have been introduced.
In particular we assume that: $\eta^{\pi^\pm}_d= \eta^{\pi^\pm}_u$, $\eta^{K^\pm}_s= \eta^{K^0}_s$,
$\eta^{e^\pm}_d= \eta^{e^\pm}_u$, $\eta^{\Lambda(\bar \Lambda)}_d= \eta^{\Lambda(\bar\Lambda)}_u$
and so on.
Within the SM, a top, with a mass above $Wb$ threshold, is predicted
to have a decay width dominated by the two-body process: $t \rightarrow W b\,\,$
($Br\simeq 0.998$)~\cite{PDG}. Then a $t \bar t$ pair will mainly decay into $Wb\,Wb$.
As we are interested to study the CKM-suppressed top quark decay $t \rightarrow s W$,
we will look for $t \bar t \rightarrow Ws\,Wb$ and $t \bar t \rightarrow Ws\,Ws$
that are available with a probability of $2(1-$Br$_{t \to Ws})$Br$_{t \to Ws}\simeq2$Br$_{t \to Ws}$
and Br$^2_{t \to Ws}$ respectively.
This decay modes can be classified according to their topological final states as follow:
$i)$ dilepton mode, where both $W$ decays are leptonic,
with 2 jets arising from the quark hadronization ($b$ or $s$) and missing transverse energy ($\not\!\!E_{\rm T}$)
coming from the undetected neutrinos;
$ii)$ Lepton+jets mode, where one $W$ decays leptonically and the other one
into quarks, with 4 jets and $\not\!\!E_{\rm T}$;
$iii)$ All jets mode, where both the $W$'s decay into quarks with 6
jets and no associated $\not\!\!E_{\rm T}$.
In our analysis, we searched for top dilepton decay candidates without considering $\tau$ leptons
in the final state. Events were asked to have two high-$p_T$ leptons, and a reconstructed
dilepton invariant mass $M_{\ell^+\ell^-}$ outside the $Z^0$ mass window.
The jet-tagging requirements were: one tagged $b-$jet, vetoing the presence of an extra $b-$jet and also
vetoing the presence of a tagged $c-$jet. The discrimination of $s-$jets
from other light-quark-jets have been achieved by using the large flavour tagging technique.
By assuming an ILC integrated luminosity of $1$ $ab^{-1}$, and two running experiments at the same time,
we found that a branching ratio sensitivity up to $10^{-3}$ for the process $t \to sW$ is achievable.
This makes possible to search for physics beyond the SM in the top decay channel $t \to s\,W$.

\section{Conclusions and perspectives}

Rare decays of the top quark can be interesting probes of new physics. Within the SM one has
$BR(t\to s+W) \simeq 1.5 \times 10^{-3}$. When one allows RH top currents the CKM element gets modified
in such a manner that the B.R for the decay $t \to s W$ gets supressed.
In the MSSM, we can get an enhancement of order 50\%, which
may help to make it detectable at ILC. Assuming a center of mass energy of $\sqrt s=$ $500$ GeV, and a total
integrated luminosity of $1$ ab$^{-1}$ at ILC, it will be possible to reach sensitivity for
BR's up to $10^{-3}$. Thus we can summarize our results as follows:

\begin{enumerate}
\item Rare decays of the top quark can be interesting
probes of new physics.
\item $B.R.(t\to s+W) \simeq 1.5 \times 10^{-3}$ is reached in the SM.
\item Assuming a RH top current results in a suppression of $V_{ts}$ and therefore in $BR(t \to s W)$.
\item In the MSSM, we can get an enhancement of order 50 percent,
  which may help to make it detectable at ILC.

\item Thus, the future ILC will allow us to complete our understanding of top quark physics,
by measuring the CKM suppressed decay $t \to s W$.
\end{enumerate}

\section{Acknowledgments}
This work was supported by CONACYT-SNI (Mexico) and PAPIIT project No. IN113206.



\end{document}